\documentclass[prc,twocolumn,showpacs,preprintnumbers]{revtex4}
\usepackage{bm}
\usepackage{amssymb}
\usepackage{amsmath}
\usepackage{graphicx}

\begin{document}
\title{Molecular Dynamics Simulation of Shear Moduli for Coulomb Crystals}
\author{C. J. Horowitz}\email{horowit@indiana.edu} 
\author{J. Hughto}\email{jhughto@astro.indiana.edu}
\affiliation{Department of Physics and Nuclear Theory Center,
             Indiana University, Bloomington, IN 47405}

\date{\today}
\begin{abstract}
Torsional (shear) oscillations of neutron stars may have been observed in quasiperiodic oscillations of Magnetar Giant Flares. The frequencies of these modes depend on the shear modulus of neutron star crust.   We calculate the shear modulus of Coulomb crystals from molecular dynamics simulations.  We find that electron screening reduces the shear modulus by about 10\% compared to previous Ogata et al.   results.  Our MD simulations can be extended to calculate the effects of impurities and or polycrystalline structures on the shear modulus.
\end{abstract}
\smallskip
\pacs{62.20.de, 
97.60.Jd, 
52.27.Gr 
}
\maketitle

Recently quasi-periodic oscillations (QPOs) have been observed in the tails of Magentar Giant Flares \cite{qpo1}\cite{qpo2}.  These flares are extremely energetic $\gamma$-ray bursts from very strongly magnetized neutron stars.   The QPOs have been interpreted as shear oscillations of the crust \cite{QPOinterp},\cite{QPOinterp2}.  If this interpretation is correct, the QPO frequencies could provide detailed information on neutron stars and their crusts \cite{QPOinterp3}.  The frequencies of shear modes depend on shear moduli of neutron star crust which is a Coulomb solid.  Ogata et al. \cite{ogata} have calculated shear moduli using Monte Carlo simulations.  In this paper we improve on the Ogata et al. results by presenting molecular dynamics simulations with much better statistics and we include the effects of electron screening.

Our results for shear moduli are preparation for later molecular dynamics calculations of the breaking strain of neutron star crust \cite{future}.  This breaking strain determines the maximum height of neutron star ``mountains" before they collapse under their own weight.  Mountains on rapidly rotating neutron stars may efficiently radiate gravitational waves \cite{crustmonster}.  These waves could limit the spin frequencies of accreting stars and may be detectable with large scale interferometers \cite{watts}.  In addition the breaking strain may be important for crust breaking models of Magnetar Giant Flares \cite{duncan}.

In the crust of a neutron star electrons form a very degenerate relativistic gas.  The ions are completely pressure ionized and have Coulomb interactions that are screened at large distances by the slightly polarizable electron gas.  The interaction potential between two ions, $v(r)$, is assumed to be \cite{screening},
\begin{equation}
v(r)=\frac{Z^2 e^2}{r}{\rm e}^{-r/\lambda_e}\, ,
\end{equation}
where the ions have chage $Z$, $r$ is the distance between them, and the electron screening length $\lambda_e$ is 
\begin{equation}
\lambda_e=\frac{\pi^{1/2}}{2e(3\pi^2n_e)^{1/3}}
\end{equation}
with $n_e$ the electron density.  The total potential energy is $V_{tot}=\sum_{i<j}v(r_{ij})$.  Charge neutrality ensures that $n_e=Zn$ where $n$ is the ion density.  The ions are assumed to form a classical one component plasma (OCP) that can be characterized by the Coulomb parameter $\Gamma$,
\begin{equation}
\Gamma=\frac{Z^2e^2}{a T}\, .
\end{equation}
This parameter is the ratio of a typical Coulomb to thermal energy and the ion sphere radius $a=[3/(4\pi n)]^{1/3}$ characterizes the separation between ions.  The OCP is expected to freeze for $\Gamma\ge 175$.

To calculate shear moduli, we follow the procedure of Ogata et al. \cite{ogata}.  The change in free energy with deformation $\delta F$ can be expressed in terms of elastic constants $c_{11}$, $c_{12}$ and $c_{44}$,
\begin{equation}
\delta F = \frac{1}{2} (c_{11}-c_{12}) u_{ii}^2 + c_{44} u_{ik}u_{ki}\ \ \ (i\neq k)\, ,
\end{equation}
and $u_{ik}$ describes the strain.

Under a deformation, the coordinates $r_k$ of an ion get mapped to $r'_i$,
\begin{equation}
r'_i=\sum_{k=1}^3 (\delta_{ik}+u_{ik}) r_k\, .
\end{equation}
We consider six deformations $D_i$ ($i=1...6$) that conserve the volume to order $\epsilon^2$.
\begin{equation}
D_1:\ \ \ u_{xx} = \epsilon +\frac{3}{4}\epsilon^2\, ,\ \ \ u_{yy}=u_{zz}=-\frac{\epsilon}{2}
\end{equation}
\begin{equation}
D_2:\ \ \ u_{yy} = \epsilon +\frac{3}{4}\epsilon^2\, ,\ \ \ u_{xx}=u_{zz}=-\frac{\epsilon}{2}
\end{equation}
\begin{equation}
D_3:\ \ \ u_{zz} = \epsilon +\frac{3}{4}\epsilon^2\, ,\ \ \ u_{xx}=u_{yy}=-\frac{\epsilon}{2}
\end{equation}
\begin{equation}
D_4:\ \ \ u_{xy}=u_{yx}=\frac{\epsilon}{2}\, ,\ \ \ u_{zz}=\frac{\epsilon^2}{4}
\end{equation}
\begin{equation}
D_5:\ \ \ u_{yz}=u_{zy}=\frac{\epsilon}{2}\, ,\ \ \ u_{xx}=\frac{\epsilon^2}{4}
\end{equation}
\begin{equation}
D_6:\ \ \ u_{zx}=u_{xz}=\frac{\epsilon}{2}\, ,\ \ \ u_{yy}=\frac{\epsilon^2}{4}
\end{equation}
For each deformation $D_m$ we calculate a corresponding expectation value $f_m$ ($m=1...6$),
\begin{equation}
f_m=\frac{1}{V}\Bigl\{\bigl\langle \frac{d^2V_{tot}}{d\epsilon^2}\bigr\rangle -\frac{1}{T}\bigl[\langle \bigl(\frac{dV_{tot}}{d\epsilon}\bigr)^2\bigr\rangle - \bigl\langle \frac{dV_{tot}}{d\epsilon}\bigr\rangle^2 \bigr]\Bigr\},
\label{fm}
\end{equation}
where $V$ is the system volume.  At zero temperature, this reduces to $f_m=(d^2V_{tot}/d\epsilon^2)/V$.

For a body centered cubic crystal one has \cite{ogata},
\begin{equation}
f_1=f_2=f_3=3b_{11}=3(c_{11}-c_{12}) 
\end{equation}
and
\begin{equation}
f_4=f_5=f_6=c_{44}.
\end{equation}
Here $c_{11}$, $c_{12}$, and $c_{44}$ are elastic constants.  In practice we calculate all six $f_m$ independently and average to determine $b_{11}$ and $c_{44}$.  The angle averaged shear modulus is \cite{ogata},
\begin{equation}
\mu_{\rm eff}=(2b_{11}+3c_{44})/5.
\end{equation}
If neutron star crust involves many crystal domains of random orientation, then $\mu_{\rm eff}$ is the appropriate elastic constant to determine the speed of shear waves.

\begin{figure}[ht]
\begin{center}
\includegraphics[width=3.75in,angle=0,clip=true] {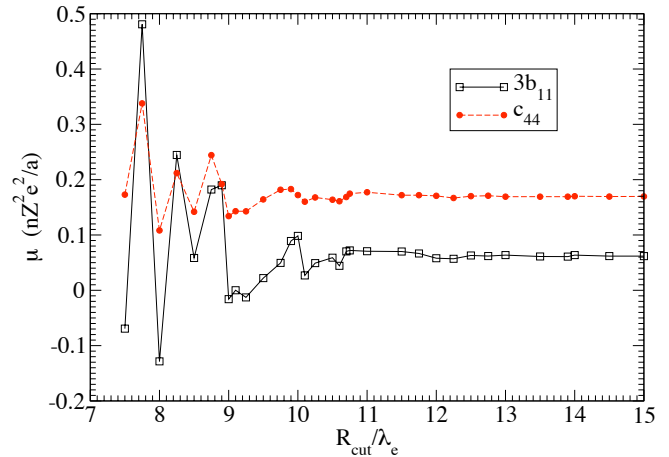}
\caption{(Color on line) Elastic constants $3b_{11}$ and $c_{44}$ versus cutoff distance $R_{\rm cut}$ for a perfect bcc lattice at zero temperature.  The cutoff distance is in units of the electron screening length $\lambda_e$.}
\label{Fig1}
\end{center}
\end{figure}

The shear modulus is sensitive to the very long range tails of the interactions.  To study this we cut off the potential at a large distance $R_{\rm cut}$,
\begin{equation}
v(r)\rightarrow v_{\rm cut}(r)=[v(r)-v(R_{\rm cut})]\Theta(R_{\rm cut} - r).
\end{equation}
We have subtracted a constant so that $v_{\rm cut}(r)$ is continuous at $r=R_{\rm cut}$.  In Fig. \ref{Fig1} we plot the elastic constants $b_{11}$ and $c_{44}$ versus $R_{\rm cut}$ for a perfect bcc lattice at zero temperature.  This figure was calculated assuming $Z=29.4$.  We note that the ratio of $\lambda_e$ to $a$ is $\lambda_e/a=5.41/Z^{1/3}$ independent of density.  We see that one must go to very large $R_{\rm cut}> 12 \lambda_e$ to calculate both $b_{11}$ and $c_{44}$ accurately.  For $R_{\rm cut}\rightarrow \infty$ we have $\mu_{\rm eff}=0.1108 (nZ^2e^2/a)$.  This is 8\% smaller than the value $\mu_{\rm eff}=0.1194 (nZ^2e^2/a)$ that Ogata et al. \cite{ogata}, calculate in the limit $\lambda_e\rightarrow \infty$.  We conclude that electron screening , neglected in ref. \cite{ogata}, reduces $\mu_{\rm eff}$ by about 10\%.

We now describe our MD simulations at finite temperatures.  For simplicity we work at a density $n=7.18\times 10^{-5}$ fm$^{-3}$ and $Z=29.4$.  Our results can be scaled to other densities at a given value of $\Gamma$.  Our results can also be approximately scaled to other values of $Z$, at fixed $\Gamma$.   This is because, although the ratio $\lambda_e/a$ changes with $Z$, this change in screening has only a small effect on the shear modulus.  We evolve the system with the velocity Verlet algorithm  \cite{verlet} using a time step $\delta t=25$ fm/c.  Starting from $T=0$ and a perfect bcc lattice we increase the temperature to $T=0.1$ MeV and evolve the system for typically 100000 MD steps ($2.5\times 10^6$ fm/c) to reach thermal equilibrium.  Next we evolve for a further 250000 MD steps ($6.25\times 10^6$ fm/c) storing configurations for later calculations of elastic constants.  The temperature is then raised by of order 0.1 MeV and the process repeated.  We keep the system at a fixed temperature (approximately) by periodically rescaling the velocities.  These MD simulations are done in an undistorted cubic box using periodic boundary conditions. 

We calculate $f_m$ by averaging over 1000 configurations, each separated by 250 MD steps (6250 fm/c).  To minimize finite size effects we calculate $V_{\rm tot}$ by summing over all 27 nearest periodic images.  Thus ion $i$ is assumed to interact not only with ion $j$ at its original position but also with 26 more images of $j$ where the x, y, and z coordinates are independently shifted by 0, $+l$, or $-l$, with $l$ the box size.  The derivatives in Eq. \ref{fm} are approximated using a five point numerical formula.  We note that the MD trajectories have been calculated using periodic distances (involving only the single nearest periodic image of a given ion) to save time, while the derivatives have been calculated by summing over 27 images to minimize finite size effects.  

Table \ref{tableone} presents results for simulations using $N=3456$ ions and no cutoff $R_{\rm cut}=\infty$.  Statistical errors only are indicated in parentheses.  We caution that $b_{11}$ may have significant errors from finite size and other systematic effects.  Indeed Fig. \ref{Fig1} suggests that finite size effects could be large for this small system.  However $b_{11}$ only makes a small contribution to $\mu_{\rm eff}$.  Therefore $\mu_{\rm eff}$ in Table \ref{tableone} may be more accurate.  We fit the values of $\mu_{\rm eff}$ in Table \ref{tableone} with a simple analytic formula that is valid for all $\Gamma\ge 175$,
\begin{equation}
\mu_{\rm eff} \approx (0.1106 -\frac{28.7}{\Gamma^{1.3}})(n\frac{Z^2e^2}{a}).
\label{mufit}
\end{equation}
This fit has an error $\le 2$\%.

\begin{table}
\caption{Shear Moduli for MD simulations with $N=3456$ ions.} 
\begin{tabular}{llll}
$\Gamma$ & $b_{11}$ ($nZ^2 e^2/a$) & $c_{44}$ ($nZ^2e^2/a$) & $\mu_{\rm eff}$ ($nZ^2e^2/a$)\\
\toprule
$\infty$ & 0.0220 & 0.1699 & 0.1107 \\
834 & 0.0209(2) & 0.1617(3) & 0.1054(2)\\
417 & 0.0194(2) & 0.1517(3) & 0.0988(2) \\
278 & 0.0202(4) & 0.1410(5) & 0.0927(3) \\
200 & 0.0154(5) & 0.1253(10) & 0.0813(6) \\
175 & 0.0158(8) & 0.1152(10) & 0.0755(6)\\
\end{tabular} 
\label{tableone}
\end{table}

\begin{figure}[ht]
\begin{center}
\includegraphics[width=3.75in,angle=0,clip=true] {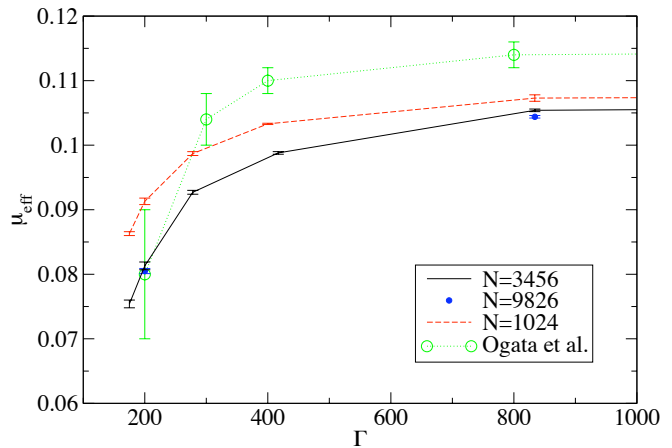}
\caption{(Color on line) Angle averaged shear modulus $\mu_{\rm eff}$ versus Coulomb parameter $\Gamma$ for MD simulations involving $N=1024$, 3456, and 9826 ions.  Also shown are Monte Carlo results from Ogata et al. \cite{ogata} that omit electron screening.}
\label{Fig2}
\end{center}
\end{figure}

To study finite size effects we have performed additional simulations with larger systems.  Table \ref{tabletwo} presents results for simulations with $N=9826$ ions using a cutoff $R_{\rm cut}=13.9\lambda_e$.  For this larger system and for finite $\Gamma$, $\mu_{\rm eff}$ is about 1\% smaller in Table \ref{tabletwo} than in Table \ref{tableone}.  Therefore we estimate finite size effects in Table \ref{tabletwo} to be of order 1\%.  Figure \ref{Fig2} plots these results for $\mu_{\rm eff}$ and in addition shows results for very small simulations with $N=1024$ ions, where finite size effects are large.
Finally, Fig. \ref{Fig2} also shows the Monte Carlo results of Ogata et al. \cite{ogata}.  These results are about 10\% larger than our results at large $\Gamma$ and have much larger statistical errors.

Ogata et al. neglect electron screening $\lambda_e\rightarrow \infty$.  At zero temperature we have performed calculations for larger values of $\lambda_e$ and extrapolated to $\lambda_e\rightarrow\infty$.  Note that we can not directly calculate for $\lambda_e=\infty$.  Our extrapolated results are consistent with Ogata et al.  Therefore we conclude that electron screening reduces $\mu_{\rm eff}$ by about 10\%.  The speed of shear waves is proportional to the square root of $\mu_{\rm eff}$.  Therefore electron screening reduces the shear speed by about 5\%.  This will slightly lower the frequency of torsional oscillations of neutron star crusts.

\begin{table}
\caption{Shear Moduli for MD simulations with $N=9826$ ions using a cutoff $R_{\rm cut}=13.9\lambda_e$.} 
\begin{tabular}{llll}
$\Gamma$ & $b_{11}$ ($nZ^2 e^2/a$) & $c_{44}$ ($nZ^2e^2/a$) & $\mu_{\rm eff}$ ($nZ^2e^2/a$)\\
\toprule
$\infty$ & 0.0212 & 0.1700 & 0.1105 \\
834 & 0.0208(2) & 0.1602(2) & 0.1045(1)\\
200 & 0.0177(5) & 0.1224(6) & 0.0805(4) \\
\end{tabular} 
\label{tabletwo}
\end{table}

In future work, we will study the impact of impurities on $\mu_{\rm eff}$ by explicitly including them in our MD simulations \cite{phasesep}.  We expect impurities to lower the shear modulus because they reduce the uniformity of the crystal.  We will also study the effect of polycrystalline structure on $\mu_{\rm eff}$ with larger scale MD simulations that include multiple crystal domains.  These multiple domains could also lead to a lower effective shear modulus.  Finally, we will calculate the breaking strain by slowly deforming the simulation volume and calculating the resulting stress.  The breaking strain is important for the maximum height of mountains on neutron stars that could be important for gravitational wave radiation.  In addition, the breaking strain is important for star ``quakes" that may trigger Magnetar Giant Flares.

In conclusion, we have calculated the shear modulus of a Coulomb plasma using MD simulations.  The shear modulus is important for the frequencies of torsional oscillations of neutron star crusts.  Our results for the angle averaged shear modulus $\mu_{\rm eff}$ are,
$\mu_{\rm eff}\approx (0.1106 - 28.7/\Gamma^{1.3}) (nZ^2e^2/a)$.
Here $n$ is the ion density, $Z$ the ion charge, and $a$ the ion sphere radius, $a=(3/4\pi n)^{1/3}$.  This formula is accurate to about 2\% and valid for Coulomb parameter $\Gamma\ge 175$.  Our results are about 10\% smaller than Ogata et al because we include electron screening.

We thank Don Berry, Kai Kadau, and Andrew Steiner for helpful discussions.  This work was supported in part by DOE grant DE-FG02-87ER40365.

\vfill\eject


\begin{thebibliography}{99} 
\bibitem{qpo1}G. Isreal et al., ApJ. {\bf 628} (2005) L53.
\bibitem{qpo2} Tod E. Strohmayer and Anna L. Watts, ApJ. {\bf 632} (2005) L111.
\bibitem{QPOinterp} T. Strohmayer, S. Ogata, H. Iyetomi, S. Ichimaru, and H. M. Van Horn, ApJ {\bf 375} (1991) 679.
\bibitem{QPOinterp2} Anthony L. Piro, ApJ. {\bf 634} (2005) L153.
\bibitem{QPOinterp3} Lars Samuelsson and Nils Andersson, Mon. Not. Roy. Astron. Soc. {\bf 374} (2007) 256. 
\bibitem{ogata} Shuji Ogata and Setsuo Ichimaru, Phys. Rev. A {\bf 42} (1990) 4867.
\bibitem{future} C. J. Horowitz and K. Kadau, to be published.
\bibitem{crustmonster} G. Ushomirsky, C. Cutler, and L. Bildsten, MNRAS {\bf 319} (2000) 902.
\bibitem{watts} A. L. Watts, B. Krishnan, L. Bildsten, and B. F. Schutz, MNRAS {\bf 389} (2008) 839.
\bibitem{duncan} C. Thompson and R. C. Duncan, ApJ. {\bf 561} (2001) 980.
\bibitem{screening}  A. L. Fetter and J. D. Walecka, Quantum Theory of Many Body Systems (McGraw-Hill, New York,1971), p. 175.
\bibitem{phasesep} C. J. Horowitz, D. K. Berry, and E. F. Brown, PRE {\bf 75} (2007) 066101.
\bibitem{verlet} L. Verlet, Phys. Rev. {\bf 159}, 98 (1967).
F. Ercolessi, A Molecular Dynamics Primer, available from http://www.sissa.it/furio/ (1997).









\end{thebibliography}
\end{document}